\documentclass[11pt]{article}
\begin{document}

\begin{center}
\begin{Large}\textbf{A Note on the Relation Between Balenzela's Algorithm for Two-Filter Formula for Smoothing and Information Filter}\end{Large}\\

\vspace{10mm}
{\large Genshiro Kitagawa}\\[2mm]
The Institute of Statistical Mathmatics \\
and\\
The Graduate University for Advanced Study

\vspace{3mm}
{\today}
\end{center}

\noindent{\bf Abstract.}

Recent paper by Balenzuela et al.\cite{BWRN 2022} presented an exact algorithm for computing the posterior distribution of current and future observations given the current state, $p(x_n|y_n,\ldots ,y_N)$, which is required when computing fixed-interval smoother of the state by a two-filter formula. In this note, it will be shown that their algorithm is equivalent to the backward filter obtained by applying an information filter to the reverse state-space model. Although their algorithm is proposed for complex Gaussian mixture distribution models, in this note, we consider the case of simple state-space models with respect to filter computation.

\vspace{2mm}

\section{Introduction}

State-space models are a versatile analytical tool for time series analysis. In the use of state-space models, the estimation of states is essential, since various important prolems in time series anlysis such as the estimation of model parameters, prediction of time series, decomposition to several components, and treatment of outliers and missing values, etc., can be performed through the estimated state. 

In particular, in nonstationary time series analysis, it is used in various forms such as trend estimation, seasonal adjustment and component decomposition.
In these tasks, fixed-interval smoothing algorithms are used to obtain the posterior distribution of the state under the condition that all data are given. Even when the noise distribution is represented by a mixed normal distribution, the predicted and filter distributions can be obtained by a Gaussian-sum filter that extends the Kalman filter, but the formula for a fixed interval smoothed distribution requires division by the predicted distribution, so it can not be directly extended to a state-space model with a Gaussian-mixture noise distribution. In this case, the two-filter formula is utilized for obtaining smoothed distribution of the state. Balenzuela et al. proposed a method to calculate an exact two-filter algorithm for a state-space model with stochastic jump process. In this note, we show that their algorithm is consistent with the application of the general information filter to the inverse model.

The plan of the paper is as follows. In Section 2, fixed-interval for the standard state-space model and two-filter algorithm for smoothing are reviewed. In Section 3, an algorithm for obtaining $p(y_{n:N}|x_n)$ is otained by applying the information filter to thebackward state-space model associated with the orginal state-space model.
In Section 4, it will be shown that the presented algorithm is equivalent to the algorithm presented in Balenzuela et al.\cite{BWRN 2022}.

\section{Linear-Gaussian State-Space Model and the Kalman Smoother}

\subsection{A brief review of the Kalman filter and the smoother}

Assume that the state-space model is linear and is given by
\begin{eqnarray}
  x_n &=& F_n x_{n-1} + G_n v_n \nonumber \\
  y_n &=& H_n x_n + w_n, \label{Eq_SSM}
\end{eqnarray}
where $y_n$ is a univariate time series, $x_n$ is $d_x$-dimensional state vector, $F_n$, $G_n$ and $H_n$ are $d_x \times d_x$, $d_x \times d_v$ and $1 \times d_x$ dimensional matrices, respectively.  
$v_n$ and $w_n$ are Gaussian white noise such that $v_n \sim N(0,Q_n)$ and 
$w_n \sim N(0,R_n)$, respectively.
It is well known that if the initial
state density is Gaussian, $x_{0|0} \sim N(x_{0|0},V_{0|0})$, 
then the conditional density $p(x_n|Y_m)$ is also
Gaussian and that the mean $x_{n|m}$ and the variance-covariance matrix $V_{n|m}$ can be obtained by the Kalman filter and the fixed interval smoothing algorithms (Anderson and Moore (1979)).

The Kalman filter consists of the following sequential computations for $n=1,\ldots ,N$:

\vspace{2mm}
\textbf{One-step ahead prediction}
\begin{eqnarray}
x_{n|n-1} &=& F_n x_{n-1|n-1}, \nonumber \\
V_{n|n-l} &=& F_n V_{n-1|n-1} F_n^T + G_nQ_nG_n^T.
\end{eqnarray}

\textbf{Filter}
\begin{eqnarray}
K_n  &=& V_{n|n-1}H_n^T(H_nV_{n|n-1}H_n^T + R_n)^{-1}, \nonumber \\
x_{n|n} &=& x_{n|n-1} + K_n(y_n - H_n x_{n|n-1}), \\
V_{n|n} &=& (I-K_n H_n)V_{n|n-1}. \nonumber
\end{eqnarray}
Using these estimates, the smoothed density of the state $x_n$ given the data $Y_N$ is obtained by the following backward recursion for $n=N-1,\ldots ,1$:

\vspace{2mm}
\textbf{Fixed interval smoothing algorithm}
\begin{eqnarray}
A_n     &=& V_{n|n} F_n^T V_{n+1|n}^{-1}, \nonumber \\
x_{n|N} &=& x_{n|n} + A_n(x_{n+1|N} - x_{n+1|n}), \\
V_{n|N} &=& V_{n|n} + A_n (V_{n+1|N} - V_{n+1|n})A_n^T. \nonumber
\end{eqnarray}
Note that the initial values for this recursion, $x_{N|N}$ and $V_{N|N}$, are obtained by the Kalman filter.

\subsection{Two-filter formula for the linear Gaussian state-space model}

In this subsection, we consider the two-filter smoothing formula for the linear Gaussian state-space model. Note that it is not essential to use the two-filter formula for the linear Gaussian state-space model, since the fixed-interval smoothing algorithm can be applied to it. However, since the Gaussian-sum smoothing algorithm is obtained by a weighted linear sum of the smoothing distributions obtained by the two-filter formula, its extension to the Gaussian-mixture noise state-space model is straightforward, except for the component reduction algorithm and evaluation criteria for the reduced component model.

In the case of the linear Gaussian state-space model, the contitional density of the 
state $x_n$ given the observations $Y_N$ is Gaussian and can be expressed as follows:
\begin{eqnarray}
  p(x_n|Y_N) &\propto& p(Y_{n:N}|x_n) p(x_n|Y_{n-1}) \nonumber \\
             &=& \varphi (Y_{n:N}|x_n) \varphi (x_n|Y_{n-1}) = \varphi (x_n|Y_N) ,
\end{eqnarray}
where $\varphi$ denotes a Gaussian density function.
If we assume that $\varphi (x_n|Y_{n-1})\sim N(x_{n|n-1},V_{n|n-1})$ and 
$\varphi (x_n|Y_{n:N})\sim N(z_{n|n},U_{n|n})$, then the smoothed density is also a Gaussin density, $\varphi (x_n|Y_N)\sim N(x_{n|N},V_{n|N})$, and its mean and the variance-covariance matrix are obtained by
\begin{eqnarray}
   J_n     &=& V_{n|n-1}\left(V_{n|n-1} + U_{n|n}\right)^{-1}, \nonumber \\
   x_{n|N} &=& x_{n|n-1} + J_n (z_{n|n} - x_{n|n-1}), \label{Eq_Two-filter-formula}\\
   V_{n|N} &=& (I - J_n) V_{n|n-1}.  \nonumber
\end{eqnarray}

Care should be taken regarding the method of calculating $z_{n|n}$ and $U_{n|n}$.
If the the varaiance-covariance matrix at the end point, $U_{N|N}$, is full rank,
$z_{n|n}$ and $U_{n|n}$ can be obtained by using the backward (reverse) state-space model:
\begin{eqnarray}
  z_n &=& F_{n+1}^{-1}z_{n+1} - F_{n+1}^{-1}G_{n+1}v_{n+1}^{(B)}, \nonumber \\
  y_n &=& H_n z_n + w_n^{(B)},  \label{Eq_Backward_SSM}
\end{eqnarray}
where $v_{n+1}^{(B)} \sim N(0,Q_n)$ and $w_n^{(B)} \sim N(0,R_n)$.
Therefore, we can apply the same Kalman filter by replacing $F_n$ and $G_n$ by
$\bar{F}_{n+1} = F_{n+1}^{-1}$ and $\bar{G}_{n+1} = -F_{n+1}^{-1}G_{n+1}$. Namely the backward filter is given by:

\vspace{2mm}
\textbf{Backward one-step ahead prediction:}
\begin{eqnarray}
z_{n|n+1} &=& \bar{F}_{n+1} z_{n+1|n+1}, \nonumber \\
U_{n|n+l} &=& \bar{F}_{n+1} U_{n+1|n+1} \bar{F}_{n+1}^T + \bar{G}_{n+1}Q_{n+1}\bar{G}_{n+1}^T.\nonumber
\end{eqnarray}

\textbf{Backward filter:}
\begin{eqnarray}
\bar{K}_n  &=& U_{n|n+1}H_n^T(H_nU_{n|n+1}H_n^T + R_n)^{-1}, \nonumber \\
z_{n|n} &=& z_{n|n+1} + \bar{K}_n(y_n - H_n z_{n|n+1}), \\
U_{n|n} &=& (I- \bar{K}_n H_n)U_{n|n+1}. \nonumber
\end{eqnarray}

Note that there are two problems with this backward filter:
\begin{enumerate}
\item If some of the eigen values of $F_{n+1}$ is less than 1, then some of the eigen values of $F_{n+1}^{-1}$ are greater than 1. Even if the eigenvalues of matrix $F_n$ are greater than 1, if $(F_n,G_n)$ is controllable and $(H_n^T,F_n^T)$ is observable, the Kalman filter is asymptotically stable and thus computable, but the numerical properties are somewhat worse.
\item Setting the initial distribution of the backward-looking filter can be problematic.If the matrix $H_N$ is not of full rank, it is not possible to obtain the initial
distribution of $x_N$ by directly inverting the observation model $y_N=H_Nx_N+w_N$. 

A practical way to define the initial distribution of $z_N$ is to put
\begin{eqnarray}
 z_{N|N} = x_{N|N},\quad U_{N|N} = V_{N|N}.
\end{eqnarray}
This initial distribution yields an approximation to the fixed interval smoothing. However, it is known that for the initial several steps,
the variances of the smoothed state are too small.

One way to mitigate this problem is to set the diagonal elements of the initial
variance-covariance matrix as $\nu^2/M$, where $\nu$ is the variance of the time 
series in one cycle, i.e., $\nu = p^{-1}\sum_{j=1}^p (y_{N-j+1}-\mu)$, 
$\mu = p^{-1}\sum_{j=1}^p y_{N-j+1}$, $p$ is the length of one cycle and $m$
is the dimension of the state. In the example in the next section, it will be
shown that very close approximation to the fixed interval smoother can be
obtained by this modification.

Kitagawa (1994) shows a method of defining the initial state vector $x_{N-m+1}$ 
rather than $x_N$ to garantee the full-rank of the initial variance-covariance matrix.
On the other hand Balenzuela et al. (2022) proposed a new smoothing algorithm based on the backward information filter.
\end{enumerate}

\section{Information Filter and Backward Information Filter}

Hereafter, we shall show a method based on the information filter (Kaminski 1971) for the backward state-space model (\ref{Eq_Backward_SSM}).
The information filter propagates information matrix (inverse of the variance-covariance matrix) instead of the variance-covariance matrix\cite{Fraser 1967}\cite{Kaminski 1971}.
Assume that the mean and the variance-covariance matrix of the conditional
distribution $p(x_n|Y_{n-1})$ and $p(x_n|Y_{n})$ are denoted as
$x_{n|n-1}$, $V_{n|n-1}$, $x_{n|n}$ and $V_{n|n}$, respectively.
Instead of the mean vectors and the variance-covariance matrices,
the (forward) information filter propagates $d_{n|n-1}$, $d_{n|n}$, $V_{n|n-1}^{-1}$ and
$V_{n|n}^{-1}$, where $d_{n|nn+1}$ and $d_{n|n}$ are defined by
\begin{eqnarray}
  d_{n|n-1} = V_{n|n-1}^{-1} x_{n|n-1}, \quad d_{n|n} = V_{n|n}^{-1} x_{n|n}.
\end{eqnarray}

The information filter for the forward state-space model (\ref{Eq_SSM}) is given by:\cite{Fraser 1967}\cite{Kaminski 1971}

\vspace{2mm}
\textbf{Information predictor}
\begin{eqnarray}
\Phi_n &=& F_n^{-T} V_{n-1|n-1}^{-1} F_n^{-1} \nonumber \\
  L_n  &=& \Phi_n G_n (Q_n^{-1} + G_n^T\Phi_n G_n)^{-1} \nonumber \\
%       &=& F_n^{-T} V_{n-1|n-1}^{-1} F_n^{-1} G_n (Q_n^{-1} + G_n^T\Phi_n G_n)^{-1} \nonumber \\
  d_{n|n-1} &=& (I- L_nG_n^T)F_n^{-T} d_{n-1|n-1} \label{Eq_Information-Predictor} \\
%       &=& F_n^{-T} d_{n-1|n-1} - F_n^{-T} V_{n-1|n-1}^{-1} F_n^{-1} G_n (Q_n^{-1} \!+ \!G_n^T\Phi G_n)^{-1}_nG_n^TF_n^{-T} d_{n-1|n-1} \nonumber \\
  V_{n|n-1}^{-1} &=& (I- L_nG_n^T) \Phi_n . \nonumber  % \\
%       &=& \Phi_n - \Phi_n G_n (Q_n^{-1} + G_n^T\Phi G_n)^{-1} G_n^T  \Phi_n \nonumber \\
%       &=& (\Phi_n^{-1} + G_n Q_n G_n^T)^{-1}  \nonumber \\
%       &=& (F_n V_{n-1|n-1} F_n^T + G_n Q_n G_n^T)^{-1}.  \nonumber 
\end{eqnarray}

\textbf{Information filter}
\begin{eqnarray}
  d_{n|n} &=& d_{n|n-1} + H_n^T R_n^{-1} y_n  \nonumber \\
  V_{n|n}^{-1} &=& V_{n|n-1}^{-1} + H_n^T R_n^{-1} H_n . \label{Eq_Information-Filter} 
\end{eqnarray}

Here we consider an application of this information filter to the backward
state-space model (\ref{Eq_Backward_SSM}).
Assume that the mean and the variance-covariance matrix of the backward 
predictor and filter are denoted as $p(x_n|Y_{n+1:N}) \sim N(z_{n|n+1},U_{n|n+1})$ and $p(x_n|Y_{n:N}) \sim N(z_{n|n},U_{n|n})$, respectively.
Further we assume that $d_{n|n+1} = U_{n|n+1}^{-1} z_{n|n+1}$ 
and $ d_{n|n} = U_{n|n}^{-1} z_{n|n}$.

Then to apply to the backward state-space model (\ref{Eq_Backward_SSM}), 
by replacing $F_n$ by $F_{n+1}^{-1}$ and $G_n$ by $-F_{n+1}^{-1}G_{n+1}$
in the forward information predictor(\ref{Eq_Information-Predictor}) and
the filter (\ref{Eq_Information-Filter}), we obtain:

\vspace{5mm}
\textbf{Backward information predictor}
\begin{eqnarray}
   L_n  &=& -F_{n+1}^{T} U_{n-1|n-1}^{-1} G_{n+1} (Q_{n+1}^{-1} + G_{n+1}^TU_{n+1|n+1}^{-1}G_{n+1})^{-1}, \nonumber \\
  d_{n|n+1} &=& (F_{n+1}^T +L_nG_{n+1}^T) d_{n+1|n+1}, \nonumber \\
  U_{n|n+1}^{-1} &=& (F_{n+1}^T + L_nG_{n+1}^T) U_{n+1|n+1}^{-1} F_{n+1}. \nonumber  
\end{eqnarray}

\textbf{Backward information filter}
\begin{eqnarray}
  d_{n|n} &=& d_{n|n+1} + H_n^T R_n^{-1} y_n,  \nonumber \\
  U_{n|n}^{-1} &=& U_{n|n+1}^{-1} + H_n^T R_n^{-1} H_n. \nonumber 
\end{eqnarray}

\textbf{[Proof]}
Substituting $F_{n+1}^{-1}$ for $F_n$, $-F_{n+1}^{-1}G_{n+1}$ for $G_n$
and $Q_{n+1}$ for $Q_n$ in equation (\ref{Eq_Information-Predictor}), we obtain:
\begin{eqnarray}
 \Phi_n &=& F_{n+1}^T U_{n+1|n+1}^{-1} F_{n+1} \nonumber \\
  L_n  &=& -\Phi_n F_{n+1}^{-1}G_{n+1} (Q_{n+1}^{-1} + G_{n+1}^TF_{n+1}^{-T}\Phi_n F_{n+1}^{-1}G_{n+1})^{-1} \nonumber \\
       &=& -F_{n+1}^{T} U_{n-1|n-1}^{-1} G_{n+1} (Q_{n+1}^{-1} + G_{n+1}^TU_{n+1|n+1}^{-1}G_{n+1})^{-1} \nonumber \\
  d_{n|n+1} &=& (I + L_nG_{n+1}^TF_{n+1}^{-T})F_{n+1}^T d_{n+1|n+1} \nonumber \\
       &=& (F_{n+1}^T +L_nG_{n+1}^T) d_{n+1|n+1} \nonumber \\
  U_{n|n+1}^{-1} &=& (I + L_nG_{n+1}^TF_{n+1}^{-T}) \Phi_n \nonumber \\
       &=& (F_{n+1}^T + L_nG_{n+1}^T) U_{n+1|n+1}^{-1} F_{n+1}.  \nonumber 
\end{eqnarray}

\vspace{2mm}
Note that in the case of backward information filter, the inverse of the
transition matrix $F_n^{-1}$ is replaced by the transpose of the matrix
$F_{n+1}^T$.

For the situation that $U_{n|n}^{-1}$ and $U_{n|n+1}^{-1}$ are not invertible
that apply for the first $m$ steps, i.e., $n=N,\ldots, N-m+1$,
we can use the following formula for two-filter smoothing that does not
contain $U_{n|n}$:

\vspace{2mm}
\textbf{Two-Filter Formula for Smoothing (1)}
\begin{eqnarray}
  V_{n|N} &=& (V_{n|n-1}^{-1} + U_{n|n}^{-1})^{-1}  \nonumber \\
  x_{n|N} &=& V_{n|N}(V_{n|n-1}^{-1}x_{n|n-1} + d_{n|n}). \nonumber 
\end{eqnarray}
Note that we can also obtain the smoothing distribution, $x_{n|N}$ and $V_{n|N}$,
from the forward filter, $x_{n|n}$ and $V_{n|n}$ and the backward information 
predictor $z_{n|n+1}$ and $U_{n|n+1}$ as follows:

\vspace{2mm}
\textbf{Two-Filter Formula for Smoothing (2)}
\begin{eqnarray}
  V_{n|N} &=& (V_{n|n}^{-1} + U_{n|n+1}^{-1})^{-1} \nonumber \\
  x_{n|N} &=& V_{n|N}(V_{n|n}^{-1}x_{n|n} + d_{n|n+1}). \nonumber
\end{eqnarray}

\noindent
\textbf{[Proof of the two-filter formula for smoothing (1)]}

From equation (\ref{Eq_Two-filter-formula}), using the matrix inversion lemma\cite{AM 1979}\cite{KG 1996}
\begin{eqnarray}
  J_n     &=& V_{n|n-1}(V_{n|n-1} + U_{n|n})^{-1} \nonumber \\
          &=& (V_{n|n-1}^{-1} + U_{n|n}^{-1})^{-1}U_{n|n}^{-1} \nonumber \\
  V_{n|N} &=& (I- J_n) V_{n|n-1}  \nonumber \\
          &=&  (V_{n|n-1}^{-1} + U_{n|n}^{-1})^{-1}(V_{n|n-1}^{-1} + U_{n|n}^{-1} - U_{n|n}^{-1})V_{n|n-1} \nonumber \\
          &=& (V_{n|n-1}^{-1} + U_{n|n}^{-1})^{-1} V_{n|n-1}^{-1}V_{n|n-1} \nonumber \\
          &=& (V_{n|n-1}^{-1} + U_{n|n}^{-1})^{-1} \nonumber \\
  x_{n|N} &=& x_{n|n-1} + J_n(z_{n|n} - x_{n|n-1}) \nonumber \\
          &=& (I-J_n)z_{n|n-1} + J_n z_{n|n} \nonumber \\
          &=& (V_{n|n-1}^{-1} + U_{n|n}^{-1})^{-1}(V_{n|n-1}^{-1}z_{n|n-1} + U_{n|n}^{-1}z_{n|n}) \nonumber \\
          &=& V_{n|N}(V_{n|n-1}^{-1}z_{n|n-1} + d_{n|n}) \nonumber
\end{eqnarray}

\textbf{[Proof of the two-filter formula for smoothing (2)]}
\begin{eqnarray}
  J_n     &=& V_{n|n}(V_{n|n} + U_{n|n+1})^{-1} \nonumber \\
          &=& (V_{n|n}^{-1} + U_{n|n+1}^{-1})^{-1}U_{n|n+1}^{-1} \nonumber \\
  V_{n|N} &=& (I- J_n) V_{n|n}  \nonumber \\
          &=& (V_{n|n}^{-1} + U_{n|n+1}^{-1})^{-1}(V_{n|n}^{-1} + U_{n|n+1}^{-1} - U_{n|n+1}^{-1})V_{n|n} \nonumber \\
          &=& (V_{n|n}^{-1} + U_{n|n+1}^{-1})^{-1} \nonumber \\
  x_{n|N} &=& x_{n|n} + J_n(z_{n|n+1} - x_{n|n}) \nonumber \\
          &=& (I-J_n)x_{n|n} + J_n z_{n|n+1}  \nonumber\\
          &=& V_{n|N}V_{n|n}^{-1}x_{n|n} + (V_{n|n}^{-1} + U_{n|n+1}^{-1})^{-1} d_{n|n+1} \nonumber \\
          &=& V_{n|N}V_{n|n}^{-1}x_{n|n} + V_{n|N} d_{n|n+1} \nonumber \\
          &=& V_{n|N}(V_{n|n}^{-1}x_{n|n} + d_{n|n+1}). \nonumber
\end{eqnarray}

\section{Relationship with the Balenzuela's Smoothing Algorithm}

In this section, it will be shown that the backward information filter obtained
in the previous section is equivalent to the smoothing algorithm proposed by Balenzuela et al.\cite{BWRN 2022}. 
Although, the smoothing algorithm was derived for jump Markov linear systems,
we consider here a simple linear Gaussian state-space model.
In the jump Markov linear model, the Guassian-sum filtering is necessary.
However, the extension to the Gaussian-sum filtering itself is rather straightforward,
except for the mixture reduction problem, for which algorithm for reducing the
number of Gaussian components and a proper criterion to find optimal 
approximation with smaller number of Gaussian components are difficult problems.

In smoothing algorithm by Balenzuela et al., for the simple state-space model, 
the mean $x_{n|N}$ and variance-covairance matrix $V_{n|N}$ of the smoothed 
distribution $p(x_n|Y_{1:N}) \sim {\mathcal N}(x_n|x_{n|N},V_{n|N})$ is 
obtained by the two filter smoothing algorithm\cite{BWRN 2022}:

\newpage
\vspace{2mm}
\textbf{Two-filter smoother}

{}\hspace{10mm}For $n=N-1,\ldots ,1$
\begin{eqnarray}
   V_{n|N} &=&  (C_n + V_{n|n}^{-1})^{-1} \nonumber \\
   x_{n|N} &=&  V_{n|N}\left( V_{n|n}^{-1}x_{n|n} - b_n \right) ,
\end{eqnarray}
where $b_n$ and $C_n$ are obtained by the following backwards information filter.

%%%%%%%%%%%%%%%%%%%%%%%%%%%%%%%%%%%%

\textbf{Backward Information Filtering}

\vspace{2mm}
Assume that the function ${\mathcal L}(x|a,b,C)$ is defined by
\begin{eqnarray}
 {\mathcal L}(x|a,b,C) &\equiv& \exp \left\{ -\frac{1}{2}(a + 2x^T b + x^TCx ) \right\} .
\end{eqnarray}

\textbf{Initialization:}
\begin{eqnarray}
 p(y_N|x_N) = {\mathcal L}(x_N|a_N,b_N,C_N)
\end{eqnarray}
where
\begin{eqnarray}
   a_N &=&  y_N r_N^{-1} y_N + \log (2\pi R_N) \nonumber \\
   b_N &=&  -H_N^T R_N^{-1} y_N \nonumber \\
   C_N &=&  H_N^T R_N^{-1} H_N . \nonumber 
\end{eqnarray}

Then the backward information filter can be performed by applying the following
algorithm for $n=N-1,\ldots 1$.
Note that, since tha algorithm of Balenzuela et al. assumes that $G=I$, in our
case $Q$ must be replaced with $G_{n+1}Q_{n+1}G_{n+1}^T$.

\vspace{2mm}
\textbf{Backward information prediction}
\begin{eqnarray}
   p(y_{n+1:N}|x_n) = {\mathcal L}(x_n|a_n,b_n,C_n) 
\end{eqnarray}
{}\hspace{30mm}where
\begin{eqnarray}
   \bar{\Phi}_n &=& (I + \bar{C}_{n+1}G_{n+1}Q_{n+1}G_{n+1}^T)^{-1} \bar{C}_{n+1} \nonumber \\
   \Psi_n &=& G_{n+1}Q_{n+1}G_{n+1}^T \bar{\Phi}_n G_{n+1}Q_{n+1}G_{n+1}^T - G_{n+1}Q_{n+1}G_{n+1}^T \nonumber \\
   \Gamma_n &=& I - G_{n+1}Q_{n+1}G_{n+1}^T \bar{\Phi}_n \nonumber \\
   a_n &=&  \bar{a}_{n+1} - \log \Gamma_n  %%%% - 2\log T_n 
          + \bar{b}_{n+1}^T \Psi_n \bar{b}_{n+1}  \nonumber \\
   b_n &=&  F_{n+1}^T \Gamma_n^T  \bar{b}_{n+1} \nonumber \\
   C_n &=&  F_{n+1}^T \bar{\Phi}_n F_{n+1} .  \nonumber 
\end{eqnarray}

\textbf{Backward information filter}
\begin{eqnarray}
   p(y_{n:N}|x_n)  = {\mathcal L}(x_n|\bar{a}_n,\bar{b}_n,\bar{C}_n) \label{Eq_Initial_for_BIF}
\end{eqnarray}
{}\hspace{30mm}where
\begin{eqnarray}
   \bar{a}_n &=&  a_n + \log 2\pi R_n  + y_n R_n^{-1} y_n  \nonumber \\
   \bar{b}_n &=&  b_n - H_n^T R_n^{-1}y_n \nonumber \\
   \bar{C}_n &=&  C_n + H_n^T R_n^{-1} H_n . \nonumber 
\end{eqnarray}

\noindent
\textbf{[Proof of the equivalence of two information filters]}

\vspace{2mm}
To show the equivalence of the backward information filter and the
Balenzuela's algorithm, it is sufficient to show that
$C_n = U_{n|n+1}^{-1}$ and $ b_n= -d_{n|n+1}$ assuming that
$\bar{C}_{n+1}^{-1} = U_{n+1|n+1}$ and $\bar{b}_{n+1} = d_{n+1|n+1}$.
$C_n$ can be expressed as follows.
\begin{eqnarray}
   C_n &=& F_{n+1}^T \bar{\Phi}_n F_{n+1}  \nonumber \\
       &=& F_{n+1}^T (I + \bar{C}_{n+1}G_{n+1}Q_{n+1}G_{n+1}^T)^{-1} \bar{C}_{n+1} F_{n+1} \nonumber \\
       &=& F_{n+1}^T (\bar{C}_{n+1}^{-1} + G_{n+1}Q_{n+1}G_{n+1}^T)^{-1} F_{n+1} . 
\end{eqnarray}
On the other hand, since $F_{n+1}^T + L_nG_{n+1}^T$ can be expressed as
\begin{eqnarray}
F_{n+1}^T + L_nG_{n+1}^T &=& 
   F_{n+1}^T  -F_{n+1}^{T} U_{n+1|n+1}^{-1} G_{n+1} (Q_{n+1}^{-1} + G_{n+1}^TU_{n+1|n+1}^{-1}G_{n+1})^{-1} G_{n+1}^T \nonumber \\
   &=&  F_{n+1}^T \Big\{ I - U_{n+1|n+1}^{-1} G_{n+1} (Q_{n+1}^{-1} + G_{n+1}^TU_{n+1|n+1}^{-1}G_{n+1})^{-1} G_{n+1}^T \Big\} \nonumber \\
   &=&  F_{n+1}^T U_{n+1|n+1}^{-1} \Big\{ U_{n+1|n+1}^{-1} - G_{n+1} (Q_{n+1}^{-1} + G_{n+1}^TU_{n+1|n+1}^{-1}G_{n+1})^{-1} G_{n+1}^T \Big\} \nonumber \\
   &=&  F_{n+1}^T U_{n+1|n+1}^{-1} ( U_{n+1|n+1} + G_{n+1}Q_{n+1} G_{n+1})^{-1} \nonumber \\
   &=&  F_{n+1}^T ( I + U_{n+1|n+1}^{-1} G_{n+1}Q_{n+1} G_{n+1})^{-1}, \nonumber 
\end{eqnarray}
$U_{n|n+1}$ is expressed as
\begin{eqnarray}
  U_{n|n+1}^{-1} &=& (F_{n+1}^T + L_nG_{n+1}^T) U_{n+1|n+1}^{-1} F_{n+1} \nonumber \\
   &=&  F_{n+1}^T ( I + U_{n+1|n+1}^{-1} G_{n+1}Q_{n+1} G_{n+1})^{-1} U_{n+1|n+1}^{-1} F_{n+1} \nonumber \\
   &=&  F_{n+1}^T ( U_{n+1|n+1} + G_{n+1}Q_{n+1} G_{n+1})^{-1} F_{n+1}.  
\end{eqnarray}
Therefore, if $\bar{C}_{n+1}^{-1} = U_{n+1|n+1}$, it can be seen that $C_n$ and
$U_{n+1|n+1}$ are identical.

Next, $b_n$ and $d_{n|n+1}$ are respectively expressed as
\begin{eqnarray}
 b_n &=& F_{n+1}^T \Gamma_n^T  \bar{b}_{n+1} \nonumber \\
     &=& F_{n+1}^T (I - G_{n+1}Q_{n+1}G_{n+1}^T \bar{\Phi}_n) \bar{b}_{n+1} \nonumber \\
     &=& F_{n+1}^T \{I - G_{n+1}Q_{n+1}G_{n+1}^T (I + \bar{C}_{n+1}G_{n+1}Q_{n+1}G_{n+1}^T)^{-1} \bar{C}_{n+1}\} \bar{b}_{n+1} \nonumber \\
     &=& F_{n+1}^T \{I - G_{n+1}Q_{n+1}G_{n+1}^T (\bar{C}_{n+1}^{-1} + G_{n+1}Q_{n+1}G_{n+1}^T)^{-1}\} \bar{b}_{n+1} \nonumber \\
     &=& F_{n+1}^T \Big[I - \Big\{\bar{C}_{n+1}^{-1} (G_{n+1}Q_{n+1}G_{n+1}^T)^{-1} + I\Big\}^{-1}\Big] \bar{b}_{n+1} \nonumber \\
     &=& F_{n+1}^T \Big[I - \Big\{(G_{n+1}Q_{n+1}G_{n+1}^T)^{-1} + \bar{C}_{n+1}\Big\}^{-1}\bar{C}_{n+1}^{-1} \Big] \bar{b}_{n+1} \nonumber \\
     &=& F_{n+1}^T \Big[\bar{C}_{n+1} - \Big\{(G_{n+1}Q_{n+1}G_{n+1}^T)^{-1} + \bar{C}_{n+1}\Big\}^{-1}\Big] \bar{C}_{n+1}^{-1} \bar{b}_{n+1} \nonumber \\
     &=& F_{n+1}^T (\bar{C}_{n+1}^{-1} + G_{n+1}Q_{n+1}G_{n+1}^T)^{-1} \bar{C}_{n+1}^{-1} \bar{b}_{n+1},  \\
  d_{n|n+1} &=& (F_{n+1}^T +L_nG_{n+1}^T) d_{n+1|n+1} \nonumber \\
   &=&  F_{n+1}^T ( I + U_{n+1|n+1}^{-1} G_{n+1}Q_{n+1} G_{n+1})^{-1} d_{n+1|n+1} \nonumber \\
   &=&  F_{n+1}^T ( U_{n+1|n+1} + G_{n+1}Q_{n+1} G_{n+1})^{-1}U_{n+1|n+1} d_{n+1|n+1} .
\end{eqnarray}
Since the filtering step is identical for both algorithms, if we set the initial 
values as equation (\ref{Eq_Initial_for_BIF}), then it can be seen that $b_n$ and
$-d_{n|n+1}$ are identical.

\end{document}